\documentstyle[preprint,prd,aps]{revtex}
\begin{document}

\title{Dirac electron in a Coulomb Field in 2+1 Dimensions}

\author{V.R. Khalilov$^1$ and Choon-Lin Ho$^2$}

\address{\small \sl
1. Department of Physics, Moscow State University\\
Moscow $119899$, Russia\\
2. Department of Physics, Tamkang University, Tamsui 25137, Taiwan}

\date{Jan 5, 1998}

\maketitle

\begin{abstract}
Exact solutions of Dirac equation in two spatial dimensions in the
 Coulomb field are obtained.  Equation which determines the so-called
critical
charge of the Coulomb field is derived and solved for a simple model.

\end{abstract}

\newpage

\section{Introduction}

Planar nonrelativistic electron systems in a uniform magnetic field are
fundamental quantum systems which have provided insights into many
novel phenomena, such as the  quantum
Hall effect
and the theory of anyons, particles obeying fractional statistics
 \cite{Prange,Wil}.
On the other hand, planar electron systems
with energy spectrum described by the Dirac Hamiltonian have also been studied
as field-theoretic models for the quantum Hall effect and
anyon theory \cite{NeS,Zeit}.
Related to these field-theoretic models are the recent
interesting studies regarding the instability
of the naive vacuum and spontaneous magnetization in (2+1)-dimensional
quantum
electrodynamics (QED), which is induced by a bare Chern-Simons
term \cite{Hoso}.
In view of these developments,  it is essential to have a better understanding
of the properties of planar Dirac particles
in the presence of external electromagnetic fields.

In this paper we would like to consider solutions of Dirac equation in two
spatial dimensions in the presence of a strong Coulomb field, and to discuss
instability of the Dirac vacuum in a
regulated strong Coulomb
field. In three space dimensions the effect of positron production by strong
Coulomb field was predicted in \cite{GZ} and were studied in
 \cite{RG,RFK,SFG,Pop,Cow,ZP,Mig,GMM,AP}.

\section{Motion of an electron in the Coulomb Field}

Let us consider a relativistic electron in two spatial dimensions in a
 Coulomb field the vector potential of which is specified as
\begin{eqnarray}
 A^0(r) = -Ze/r, \quad A^x = A^y = 0.
\label{eq1}
\end{eqnarray}
In 2+1 dimensions the Dirac matrices
may be represented in terms of the Pauli matrices. We choose the representation
$\vec{\alpha}=(-\sigma^2, \sigma^1)$ and $\beta=\sigma^3$.
Then the Dirac equation has the form ($c=\hbar=1$)
\begin{eqnarray}
(i\partial_t - H_D)\Psi = 0,
\label{eq2}
\end{eqnarray}
where
\begin{eqnarray}
  H_D = \vec{\alpha}\cdot {\bf P} + \beta m + eA^0 \equiv \sigma^1P_2
-\sigma^2P_1 + \sigma^3m + eA^0~,
\label{eq3}
\end{eqnarray}
is the Dirac Hamiltonian,
$P_{\mu} = i\partial_{\mu} - eA_{\mu}$ is the operator of generalized
momentum of electron, $m$ is the rest mass of the electron, and $e=-e_0,
e_0>0$ is its electric charge.
The conserved total angular momentum has only a single component, namely,
$J_z=L_z + S_z$, where $L_z=-i\partial/\partial\varphi$ and $S_z=\sigma^3/2$.

We shall look for solutions of (\ref{eq2}) in the form
\begin{eqnarray}
 \Psi(t,{\bf x}) = \frac{1}{\sqrt{2\pi}}\exp(-i\epsilon Et)
\psi(r, \varphi)~,
\label{e3}
\end{eqnarray}
where $\epsilon = \pm 1$, and $E>0$ is a positive quantity.  We
assume the ansatz
\begin{eqnarray}
\psi(r, \varphi) =e^{il\varphi}~
\left( \begin{array}{c}
f(r)\\
g(r)e^{i\varphi}
\end{array}\right)~,
\label{eqn6}
\end{eqnarray}
where $l$
is an integer number. The function $\psi(r,\varphi)$ is an
eigenfunction of the total angular momentum $J_z$
with eigenvalue $l+1/2$.
Substituting (\ref{e3}) and (\ref{eqn6}) in (\ref{eq2}), and taking into
account of the equations
\begin{eqnarray}
 P_x \pm iP_y = -ie^{\pm i\varphi}\left(\frac{\partial}{\partial r} \pm
\frac{i}{r}\frac{\partial}{\partial\varphi}\right)~,
\label{impul}
\end{eqnarray}
we obtain
\begin{eqnarray}
\frac{df}{dr} - \frac{l}{r}f + (\epsilon E + m + \frac{Z\alpha}{r})g = 0~,
 \nonumber \\
\frac{dg}{dr} + \frac{1+l}{r}g - (\epsilon E - m + \frac{Z\alpha}{r})f = 0~,
\label{sys}
\end{eqnarray}
where $\alpha\equiv e^2=1/137$ is the fine structure
constant.

The exact solutions and the energy eigenvalues with $\epsilon E< m$
corresponding to stationary states of the Dirac equation may be found
in full analogy with the case of three space dimensions.  We shall follow
Ref.~\cite{BLP}.
Let us look for functions $f$ and $g$ in the form
\begin{eqnarray}
 f = \sqrt{m+E}e^{-\rho/2}\rho^{\gamma-1}(Q_1 + Q_2)~, \nonumber \\
 g = \sqrt{m-E}e^{-\rho/2}\rho^{\gamma-1}(Q_1 - Q_2)~,
\label{sys1}
\end{eqnarray}
where
\begin{eqnarray}
 \rho = 2\lambda r,\quad \lambda = \sqrt{m^2-E^2},\quad \gamma =1/2 +
\sqrt{(l+1/2)^2 - (Z\alpha)^2}~.
\label{notion}
\end{eqnarray}
The value of $\gamma$ is to be found by studying the behavior
of the wave function at small $r$.
From (\ref{sys}) and (\ref{sys1}), together with the eqaulity
$(\gamma-1/2)^2 - (Z\alpha E/\lambda)^2 = (l+1/2)^2 - (Z\alpha m/\lambda)^2$,
one can derive the differential eqautions satisfied by $Q_1$ and $Q_2$.  It
turns out that the functions $Q_1$ and $Q_2$ which rendered the
solutions of (\ref{sys}) finite
at  $\rho=0$  are given in terms of the confluent hypergeometric
function $F(a, b; z)$  as
\begin{eqnarray}
 Q_1 = AF(\gamma-1/2 - (Z\alpha E/\lambda), 2\gamma; \rho)~, \nonumber \\
 Q_2 = BF(\gamma+1/2 - (Z\alpha E/\lambda), 2\gamma; \rho)~.
\label{solut}
\end{eqnarray}
The constants $A$ and $B$ are related by
\begin{eqnarray}
 B =  \frac{\gamma-1/2 - Z\alpha E/\lambda}{l+1/2 + Z\alpha m/\lambda}A~.
\label{conn}
\end{eqnarray}

The  energy eigenvalues are defined by
\begin{eqnarray}
 \gamma-\frac{1}{2} - \frac{Z\alpha E}{\lambda} = -n_r~.
\label{spect}
\end{eqnarray}
It is easy to show that the following values of the quantum number $n_r$
are allowed: $n_r =0, 1, 2,\ldots$, if $l\ge 0$, and $n_r = 1, 2, 3,\ldots$
if $l<0$.
Therefore, the electron energy spectrum in the Coulomb field (\ref{eq1}) has
the form
\begin{eqnarray}
 E = m\left[1 + \frac{(Z\alpha)^2}{(n_r + \sqrt{(l+1/2)^2-(Z\alpha)^2})^2}
\right]^{-1/2}~.
\label{spectrum}
\end{eqnarray}
It is seen that
\begin{eqnarray}
 E_0 = m \sqrt{1-(2Z\alpha)^2}~
\label{ground}
\end{eqnarray}
for $l = n_r = 0$, and  $E_0$ becomes zero at $Z\alpha=1/2$, whereas
in three spatial dimensions $E_0$ equals zero at
$Z\alpha=1$.
Thus, in two space dimensions the expression for
the electron ground state energy in the Coulomb field of a point-charge
$Z|e|$
no longer has a physical  meaning at a much lower value of
$Z\alpha=1/2$, and the corresponding solution of the Dirac
equation oscillates near the point $r\to 0$.

\section{Critical Charge}

It is known  \cite{BLP,GMM} that in three spatial dimensions the expression
for the electron ground state energy in the Coulomb field of a
point-charge  $Z|e|$
becomes purely imaginary when $Z>137$, and that
its interpretation as electron energy no longer has a physical  meaning.
To determine the
electron energy spectrum in the Coulomb field with such a charge
we need to eliminate the singularity
of the Coulomb potential of a point-charge at $r=0$ by
cutting off the Coulomb potential at small distances.  This is equivalent to
taking into account of the nucleus size.
In three space dimensions the electron energy spectrum in the Coulomb field
regulated at small distances was first considered in \cite{AP}.
With increasing $Z$ in the region $Z>137$, the electron
energy levels in  such a field were found to decrease, become negative,
and may cross the boundary of the lower energy continuum,
$E=-m$. The value of $Z|e|=Z_{\rm cr}|e|$ at which
the lowest electron energy level cross the boundary of the lower energy
continuum is called the critical charge for the
electron ground  state
 \cite{ZP,Mig,GMM}. If $Z$ continues to grow and enters the transcritical
region with  $Z>Z_{\rm cr}$, the lowest electron energy level ``sinks''
 into the lower energy  continuum, which result in a rearrangement  of the
vacuum of the QED. This rearrangement is constrained
by Pauli's exclusion principle. If the electron  ground
state at
 $Z<Z_{\rm cr}$ is vacant, two electron-positron pairs are created; if it
is half-occupied, one pair is created; and if it is
occupied, no pairs are created.
The Coulomb potential is
repulsive for the created positrons, so they go to infinity.
Hence at $Z>Z_{\rm cr}$ a quasistationary state appears in
the lower energy
continuum and the new vacuum of QED, which corresponds to
the filling of
all the electron states with $E<-m$, has the total electric charge
$2e$ \cite{ZP,Mig,GMM}. Indeed, all the electron states
with $E<-m$ (the Dirac sea) were filled at $Z<Z_{\rm cr}$, so
 electrons created by the strong Coulomb field with  $Z>Z_{\rm cr}$
cannot be described by means of a convenient wave function,
and
the notion of charged vacuum was introduced to describe these states
 \cite{RG,RFK,SFG,ZP,Mig}. In terms of the new vacuum, the density of
electric charge $\rho(r)$ is classical.  It is a function characterizing
the spatial distribution of the real electric charge appearing in the new
(charged) vacuum, while in terms of the old (uncharged) vacuum this function
should be interpreted as the probability of two electrons (with charge $2e$)
being present at a given point in space.

We would like to see how the same system behaves in two dimensions.
Let us therefore
consider the solutions and the energy eigenvalues
 corresponding to stationary states of the Dirac equation
 in the Coulomb field with  $2Z>137$  and find the
 corresponding value of  $Z_{\rm cr}$.  To find  $Z_{\rm cr}$
 it is enough to study the energy region
 near the boundary of the lower energy continuum, $-m$. We shall rewrite
 the Dirac equation,  taking account of the fact that  $\epsilon E\approx -m$.
Introducing functions  $F(r)=rf(r)$ and $G(r)=rg(r)$, and eliminating $G(r)$
from  (\ref{sys}),  we arrive at the equation for the function $F$
 near the boundary of the lower energy continuum  $-m$ in the form
\begin{eqnarray}
\frac{d^2 F(r)}{dr^2} + \left( E^2-m^2 + \frac{2\epsilon EZ\alpha}{r} +
\frac{(Z\alpha)^2 - l(l+1)}{r^2}\right)F(r) = 0~.
\label{eqnF}
\end{eqnarray}
We note that  near the boundary of the upper energy continuum for
$\epsilon E\approx m$,  the function $G(r)$ obeys the equation (\ref{eqnF})
with $F(r)$ replaced by $G(r)$.

Solution of (\ref{eqnF}), which tends to zero at $r \to \infty$,
 may be expressed by means of the Whittaker function (see, also \cite{khal})
\begin{eqnarray}
 F(r) \sim W_{\beta,i\frac{\nu}{2}}(2\lambda r)~,
\label{eqnW}
\end{eqnarray}
where
\begin{eqnarray}
\beta=\epsilon EZ\alpha/\lambda, \quad \nu=2\sqrt{(Z\alpha)^2-(l+1/2)^2},
\quad \lambda=\sqrt{m^2-E^2}~.
\label{eqnot}
\end{eqnarray}
From (7), the function $G(r)$ at $\epsilon E=-m$ can be obtained as
\begin{eqnarray}
 G(r) = \frac{1}{Z\alpha}\left((1+l)F-r \frac{dF}{dr}\right)~.
\label{eqnG}
\end{eqnarray}
Near the boundary of the upper energy continuum,
the  function $G(r)$
is given  by the Whittaker function in eq.~(\ref{eqnW}) with
$\epsilon E = m$, while
the function  $F(r)$ can be found from the relation
\begin{eqnarray}
 F(r) = \frac{1}{Z\alpha}\left(lG+r \frac{dG}{dr}\right)~.
\label{eqF}
\end{eqnarray}
Using the asymptotic representation for Whittaker
function at large $|z|$ in the form
\begin{eqnarray}
  W_{\beta,\mu}(z) \sim e^{-z/2}(z)^{\beta}~,
\label{bigWz}
\end{eqnarray}
it is seen that the bound electron state (with $|E|<m$ or  $\lambda>0$)
is localized in the plane.

Such a behavior of the wave function of the bound electron state
may be easily understood if we treat (\ref{eqnF}) as a one-dimensional
Schr$\ddot o$dinger-type equation which describes a particle with ``particle energy''
$ E^{\prime}=(E^2-m^2)/2m $
in the field of the effective potential
(in particular, for $l=0$)
$$
 U_{\rm eff}(r) = -\epsilon E Z\alpha/mr -(Z\alpha)^2/2mr^2.
$$
We note that  the effective potential is wide enough near the boundary of the
lower energy continuum  (for behavior of the effective potential in three space
dimensions see, for eg. \cite{GMM}), and that
 the effective potential in two space dimensions
 does not contain the spin electron term  $-s(s+1)\equiv -3/4$.

The solution of (\ref{eqnF}) at  $\epsilon E=-m$ can be written in terms
of the MacDonald function of imaginary order
\begin{eqnarray}
 F(r) = \sqrt{r}K_{i\nu}(\sqrt{8mZ\alpha r})~.
\label{solF}
\end{eqnarray}
The function  $G(r)$ at $\epsilon E=-m$ is determined by (\ref{eqnG}).
In the following
we shall determine the critical value $Z_{\rm cr}$ for a simple model
in which the potential $A_0(r)$ is regulated at small distances as follows:
\begin{eqnarray}
 A_0^{Z}(r) = -Ze/r, \quad r\ge R; \quad A_0^{Z}(r) = -Ze/R, r\le R.
\label{poten}
\end{eqnarray}
In the region  $r\le R$ the function $F(r)$ obeys the equation
\begin{eqnarray}
\frac{d^2 F}{dr^2} -\frac{dF}{rdr} + \left(\left(\epsilon E+
\frac{Z\alpha}{R}\right)^2 - m^2 +\frac{1-l^2}{r^2}\right)F(r) = 0~.
\label{eqnFR}
\end{eqnarray}
The solution of (\ref{eqnFR}) is
\begin{eqnarray}
 F(r) = r(A_1 J_{|l|}(\kappa r) + B_1 Y_{|l|}(\kappa r))~,
\label{solFR}
\end{eqnarray}
where
\begin{eqnarray}
 \kappa = \sqrt{\left(\epsilon E+\frac{Z\alpha}{R}\right)^2 - m^2}~,
\label{kapR}
\end{eqnarray}
and $J_n(z)$ and $Y_n(z)$ are the Bessel and the Neumann
functions of integer order $n$.

In order for the function  $F(r)$ to be finite at the point $r=0$
we need to set  $B_1=0$. To determine the energy spectrum
we need to match the solutions at the point  $r=R$:
\begin{eqnarray}
  \left(\frac{G(r)}{F(r)}\right)_{r=R-0} =
 \left(\frac{G(r)}{F(r)}\right)_{r=R+0}~.
\label{connect}
\end{eqnarray}
Taking into account of the fact
 that $R$ is much less than $1/m$, so that  $\kappa \approx Z\alpha/R$,
we obtain, for the state with $l=0$ and $\epsilon E=-m$, the following equation that determine (at fixed $R$) the critical
charge:
\begin{eqnarray}
  \frac{J_1(X)}{J_0(X)} =  \frac{1}{2X} \left(1 -
\sqrt{z}~\frac{K^{\prime}_{i\nu}(z)}{K_{i\nu}(z)}\right)~.
\label{critic}
\end{eqnarray}
Here $X=Z_{\rm cr}\alpha$,~
$\nu=\sqrt{4X^2-1},~z=\sqrt{8mRX}$, and
$K^{\prime}_{i\nu}(z)=d
K_{i\nu}(z)/dz$.
Eq. (\ref{critic})
may be solved numerically.
As we are interested only in the critical charge corresponding to the
ground state, we can consider small values of $z$.  In this case, the
Macdonald
function with imaginary order $K_{i\nu}(z)$ has the
following expansion:
\begin{eqnarray}
K_{i\nu}(z) &\to &
\sqrt{\frac{\pi}{\nu\sinh\pi\nu}}~\Biggl[\sin\left(\nu\ln\frac{2}{z}
 + \arg\Gamma(1+i\nu)\right) \\
&&+  \frac{z^2}{4\sqrt{1+\nu^2}}~\sin\left(\nu\ln\frac{2}{z}
 + \arg\Gamma(1+i\nu) + \tan^{-1}\nu \right) + \ldots \Biggr]~.
\label{Kapp}
\end{eqnarray}
Numerical solutions of eq.(\ref{critic})
give $Z_{\rm cr}\approx 84, 89$  at  $Rm = 0.02$ and $0.03$,
respectively. For comparison purpose,
we  recall
that  $Z_{\rm cr}\approx 170$ at  $Rm = 0.03$ for the analogical model in
 three space dimensions  \cite{ZP,Mig,GMM}.

Thus, the Dirac vacuum in  two space dimensions in the presence of a strong
Coulomb
field  is unstable against
electron-positron production at significantly smaller values
of the critical charge than in the case of three spatial dimensions.
Another difference between these two cases
results from the
fact that  electrons confined to a plane behave like a spinless fermion.
 So if the ground electron state at
 $Z<Z_{\rm cr}$ is vacant, one pair is created; if it is occupied, no pairs
are created.

\section{Summary}

In this paper we present the exact
solutions of the $2+1$-dimensional Dirac
equation with a Coulomb field, and
determine the crtical charge $Z_{cr}$
of a regulated Coulomb source for
which the Dirac vacuum of the system
become unstable.
At $Z>Z_{\rm cr}$ the lowest electron state  of discrete spectrum  is the
state  with $n_r\ne 0$. So if the  electron ground  state at  $Z<Z_{\rm cr}$
was vacant, then at $Z>Z_{\rm cr}$ an electron is created, together with a
hole in the lower energy
continuum.  According to Dirac  this hole is to behave
as a real positive charged particle far from the Coulomb center.
Thus, phenomena that may occur at $Z>Z_{\rm cr}$ are many-particle,
 and to describe them it is necessary to apply the quantum
field theory.
From the point of view of QED, the strong Coulomb field
with  $Z>Z_{\rm
cr}$ creates a positron and changes the vacuum in such a way that it gains the
electric charge which is exactly equal to the electron charge $e$.
The spatial distribution of the electric charge appearing in the vacuum
looks like the spatial distribution of the electron charge in the level
with $n_r=0$ in an atom with  $Z<Z_{\rm cr}$. However,
the density of the vacuum electric charge is a function
characterizing the spatial distribution of the real electric charge appearing
in the vacuum, while in the atom this function gives
 the probability density that the electron (with charge $e$) may be found at a
given point in space.

\vskip 2 truecm
\centerline{{\bf Acknowledgment}}
\medskip
This work is supported by the R.O.C. grant NSC-87-2112-M-032-003.

\vfil\eject

\end{document}